# Switchable Giant Bulk Photocurrents and Photo-spin-currents in Monolayer PT-symmetric Anti-ferromagnet MnPSe$_3$

Liang Liu, Weikang Liu, Bin Cheng, Bin Cui, Jifan Hu*

School of Physics, State key laboratory for crystal materials, Shandong University, Jinan 250100, China.

* To who should be corresponded: hujf@sdu.edu.cn

# ABSTRACT

Converting light into steady currents and spin-currents in two-dimensional (2D) platform is essential for future energy harvesting and spintronics. We show that the giant and modulable bulk photovoltaic effects (BPVEs) can be achieved in air-stable 2D antiferromagnet (AFM) monolayer MnPSe$_3$, with nonlinear photoconductance > 4000 nm·µA/V$^2$ and photo-spin-conductance > 2000 (nm·µA/V$^2$ $\hbar$/2e) in the visible spectrum. The propagation and the spin-polarizations of photocurrents can be switched via simply rotating the Néel vector. We unveil that the *PT*-symmetry, mirror symmetries, and spin-orbital-couplings are the keys for the observed sizable and controllable 2D BPVEs. All the results provide insights into the BPVEs of 2D AFM, and suggest that the layered MnPSe$_3$ is an outstanding 2D platform for energy device and photo-spintronics.

# TOC GRAPHIC

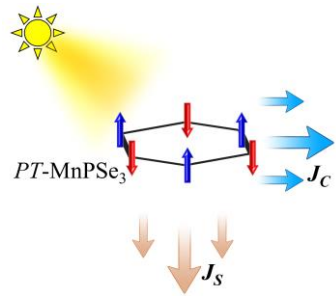

Anti-ferromagnetism enables efficient 2D solar cell in MnPSe$_3$ atomic thin layer.

Bulk photovoltaic effects (BPVEs) convert the incident light into steady currents in homogeneous crystals, sparking intense interests due to the potentials for energy harvesting, rectifications, spintronics, spectroscopes, etc. [1-9] BPVEs have been thought to be dominated by the profound shift mechanism, [10-12] in which the photocurrents are due to the continuous shifts of valence and conduction electrons in the non-centrosymmetric lattice during the photo-pumping processes. To date, the shift-type BPVEs have been realized in ferroelectrics, [3, 8, 10, 13] piezoelectrics,[14-17] and topological Weyl semimetals.[18] On the other hand, it is also possible to generate BPVEs via injection mechanism in which the net photocurrents are due to the unbalances in group velocities amongst the photo-excited carriers.[19-20] Recently, several studies have predicted that the injection-type BPVEs can be observed in two-dimensional (2D) bilayer anti-ferromagnets (AFM) $CrI_3$ [7] and $MnBi_2Te_4$,[9, 21] which possess the parity-temporal ($PT$) symmetry. $PT$-BPVEs are believed to be superior since the injection currents running in these systems are usually larger than the shift currents by several magnitude orders. More importantly, the incorporation of magnetism in BPVEs offers numerous modulation paths and opens a wealth of possibilities for opto-spintronic applications. However, the 2D materials suitable for $PT$-BPVE are very few in records. For the proposals of bilayer $PT$-AFM, the essential $PT$-symmetries stand on the long-range van der Waals (vdW) interlayer couplings, [7, 9, 21-24] and the weak vdW strengths in these interactions limit the magnitudes of photocurrents. Besides, the ubiquitous layer-slides, twists in vdW bilayers [25] might break the $PT$-symmetries and harm the BPVEs. On the other hand, the poor magnetic and air stabilities of $CrI_3$ and $MnBi_2Te_4$ cannot fulfill the demanding of realistic applications perfectly. [21-24] Hence, it is quite desired to discovery better platforms to realize $PT$-BPVEs.

$MnPSe_3$ belongs to the family of transition metal thiophosphates ($MPX_3$, with M=Mn, Fe, Co, Ni; X=S, Se), which are layered magnetic semiconductors with band gaps suitable for visible light. And they are easy to be exfoliated to 2D layers with high qualities due to the ultra-weak vdW interlayer interactions and outstanding air stabilities. [26] So, they exhibit extraordinary performances in optoelectronic responses and opto-chemistry. [27-32] However, there are notable paucity of studies investigating the BPVEs in 2D $MPX_3$, since $MPX_3$ have centrosymmetric lattice and no shift-currents are allowed. [33-35] Manganese thiophosphates are exceptional, the AFM hexagonal sublattice expanded by $Mn^{2+}$ ions break the inversion symmetry ($P$-symmetry) and the $PT$-polarization is thus hosted.[35-36] Furthermore, the anti-ferromagnetism of MnPSe$_3$ is amenable to external modulations via strains and magnetic fields.[37] Therefore, it is intriguing to see if the *PT*-AFM MnPSe$_3$ can exhibit modulable and large *PT*-BPVEs.

In this work, we predict the *PT*-BPVEs of monolayer MnPSe$_3$ induced by the illuminations of linearly polarized visible light. Based on the first-principle calculations, we show that the *PT*-polarizations in MnPSe$_3$ are stabilized by the intra-layer AFM order and the strong spin-orbital-couplings (SOC). Surprisingly, the *PT*-polarizations can perfectly align the phases of local photocurrents in (P$_2$Se$_6$)$^{4-}$ prisms, inducing large BPVEs with 2$^{nd}$ order photoconductance exceeding 4000 nm·μA/V$^2$ and photo-spin-conductance exceeding 2000 (nm·μA/V$^2$ $\hbar$/2e) in the visible spectrum. We also unveil that the 2D BPVs and the Néel vectors are intimately intertwined through the mirror symmetries, enabling abundant controlling routes for photocurrents and photo-spin-currents. It is possible to regulate the magnitudes, switch the propagations, and reversing the spin polarizations of photocurrents via rotating the Néel vectors. Hence, MnPSe$_3$ is the ideal platform to realize 2D *PT*-BPVEs, providing opportunities for high-efficient energy and controllable opto-spintronic applications.

Figure 1(a) shows the lattice of MnPSe$_3$ that consists of (P$_2$Se$_6$)$^{4-}$ prisms and two Mn-sublattices. Our density functional theory calculations indicate that the Mn$^{2+}$ ions possess local magnetic moment of 5μB, and the nearest neighbored magnetic moments are anti-parallel, consist with previous studies.[36] The anti-ferromagnetic honeycomb framework of Mn-sublattices displayed in Fig. 1(a) breaks the *P*-symmetry because the *P*-operation interchanges the two Mn-sublattices with opposite magnetic moments. Once the *P*-operation is followed by time reversal (*T*) operation which reverses all the magnetic moments, the system coincides with itself, that is, the *PT*-symmetry is preserved in the monolayer MnPSe$_3$. The lacking of *P*-symmetry indicates the MnPSe3 monolayer is a polar crystal. This kind of polarization is apparently not related to the non-centrosymmetric crystal geometry but the AFM emerged from the strong interactions between electrons, and we call this as *PT*-polarization.

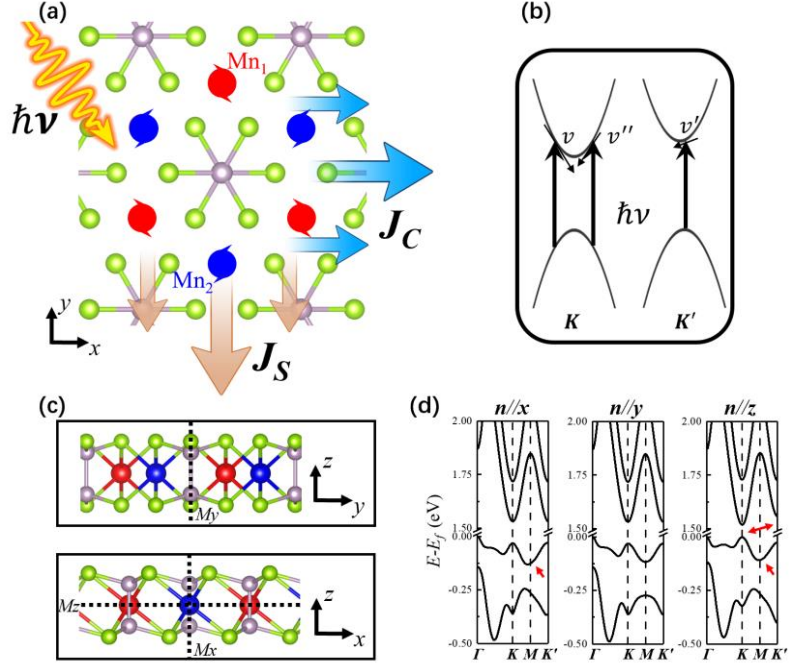

**Figure 1.** Lattice, magnetic, electronic, and photovoltaic structures of $MnPSe_3$. (a) The top view of monolayer $MnPSe_3$ lattice and the photocurrents and photo-spin-currents induced by the illuminations. The grey and green balls denote P and Se atoms. Blue and red balls denote the Mn atoms with opposite magnetic moments. (b) The typical dual valley structure in *PT*-AFM. Thick arrows denote the photo-induced hopping process and small arrows denote the group velocities of photo-excited holes. (c) The mirror reflections in $MnPSe_3$ monolayer. Dashed lines denote the mirror planes. (d) The band structure of $MnPSe_3$ with Néel vector orientation along *x*-, *y*-, and *z*-directions.

It is important to note that the *PT*-polarization found in this study is distinguished with previous reports on bilayer $CrI_3$ [7] and $MnBi_2Te_4$. [9, 21] In those systems, the *PT*-polarizations depend on the stacking order of bilayers and the long-range interlayer AFM couplings across vdW gaps. On the contrary, the *PT*-polarization in $MnPSe_3$ is induced by the strong short-range intralayer AFM exchange couplings. In addition, the intralayer crystal structure, which is based on the sturdy covalence bonds, is also more stable than the interlayer structures, which are based on the long-range vdW interactions. As a result, the monolayer $MnPSe_3$ is expected to have more robust *PT*-polarization, larger nonlinear opto-electronic couplings, and better stability.

Figure 1(b) shows the basic process of injection-type BPVs enabled by the *PT*-polarization. *K*

and $K'$ label the dual valleys with opposite momentum *i.e.*, $K' = -K$, and they are related to each other by *P*- and *T*-operation. Since both *P*- and *T*-symmetries are broken, the degenerations between $K$ and $K'$ valleys are no longer to be enforced by these two symmetries. On the other hand, the valley-polarizations are also ruled by the mirror symmetry since some mirror operations interchange the dual valleys. Given the fact that the mirror symmetries are readily switchable via simply rotating the Néel vectors, it is straight to see that the *PT*-AFM with suitable mirror symmetries possess controllable valley-polarizations, exhibiting intriguing electronic structures for modulable photoelectronic responses.

Figure 1(c) shows the two possible mirror operations in MnPSe$_3$. The first one termed as $M_y$ is the reflection about the mirror plane $M_y$ which is vertical to the basal plane and crosses the (P$_2$Se$_6$) prisms. $M_y$ interchanges the dual Mn-sublattices, and flips the magnetic moments with *x*/*z*-orientations. Therefore, MnPSe$_3$ with Néel vector along *x*/*z*-directions preserve the $M_y$-symmetry. The second mirror operation termed as $M_xM_z$ is a combination of double reflections about the one mirror perpendicular and the other vertical to the atomic plane. The Mn-sublattices do not exchange under $M_xM_z$, and the magnetic moments along *y*-directions are also preserved in $M_xM_z$. So, for the MnPSe$_3$ with Néel vector along *y*-direction, the $M_xM_z$-symmetry is preserved. More essentially, since the $M_xM_z$-operation interchanges the dual valleys while $M_y$-operation does not, the valley-polarizations in MnPSe$_3$ are prohibited in $M_xM_z$-symmetric case but allowed in $M_y$-symmetric case. Therefore, via simply rotating the magnetization orientations, we can readily obtain desired mirror-symmetry and switch the valley-polarizations in MnPSe$_3$.

Figure 1(d) shows the band structures of MnPSe$_3$ predicted by relativistic DFT, with Néel vectors along *x*-, *y*-, and *z*-directions. In all cases, the band gap emerged in the corners of Brillouin zone (BZ), either $K$ or $K'$, indicating the valley structures dominate the low-energy opto-electronic responses. For the MnPSe$_3$ with magnetic moments along *z*-axis, the asymmetries between $K$ and $K'$ valleys are obvious. The energy gaps at $K$ and $K'$ are $\Delta_K$ = 1.5196 eV and $\Delta_{K'}$ = 1.5838 eV, corresponding to the valley-polarization energy of 64.2 meV, which is a large value and consistent with previous studies.[36] For the case with Néel vectors along *x*-axis, we have $\Delta_K$ = 1.5632 eV and $\Delta_{K'}$ = 1.5635 eV, so the valley-polarization energy shrinks to 0.3 meV. The reason for this significant shrinking is that the valley-polarization is proportional to the efficiency of spin-orbit-coupling

(SOC), and the system with magnetizations along *x*-axis exhibits negligible SOC for BZ corner states. In 2D systems, the orbital moments of Bloch electrons around the BZ corners are dominated by the *z*-component, the z-orientated spins thus display the strongest SOC and valley-polarizations. For the *y*-magnetization, the valley polarization is forbidden by $M_xM_y$-symmetry and $\Delta_K = \Delta_{K'}$ =1.5634 eV.

Besides, the symmetric aspects of band structure on the inner part of BZ also show clear dependences on the magnetization orientation. For MnPSe$_3$ with Néel vectors along *x*- and *z*-directions, the asymmetries of valence bands about *M* point are shown by the red arrows marked in Fig. 1(d). The extents of band asymmetries are similar in *x*- and *z*-magnetized cases, indicating that the orbital moments around M point have comparable *x*- and *z*-components, so that the efficiencies of SOC in these states are close to each other.

BPVE is the 2$^{nd}$ order opto-electronic responses, and the photocurrents $j^\mu{}_C$ can be phenomenologically expressed as: $j_C^\mu = Re \sum_{\alpha\beta} \sigma_C^{\mu:\alpha\beta}(0;\omega,-\omega)E^\alpha(\omega)E^\beta(-\omega)$, where $\mu$, $\alpha$, and $\beta$ can be *x/y/z* directions. $\omega$ denotes the frequency of photons. $\sigma_C^{\mu:\alpha\beta}$ is the 3-ranked tensor of BPVE photoconductance, $E^{\alpha/\beta}$ denotes the $\alpha/\beta$ component of electric fields of incident light. For the system with *P*-symmetry, the *P*-operation will reverse the $j_C^\mu$ in l.h.s. but preserve the r.h.s., leading to the vanishing of BPVEs. So, the anti-ferromagnetism which produces *PT*-polarizations and lifts the *P*-symmetry in MnPSe3 is the key for the nonzero BPVEs. On the other hand, although $\sigma_C^{\mu:\alpha\beta}$ always breaks the *T*-symmetry due to the relaxation dynamics of photo-induced carriers which lacks time reciprocities, the details of how the *T*-symmetry is broke decides the magnitude and direction of *PT*-polarization, thus is also significantly relevant to *PT*-BPVEs.[19]

In the following, we focus on the BPVEs generated by the light which are linearly polarized in *x*-axis, and only the real part of $\sigma_C^{x:xx}$ and $\sigma_C^{y:xx}$ are relevant. Other directions of BPVE can be obtained by the considerations on symmetry. Microscopically, the BPVE can be evaluated by the nonlinear response theory, and the dominated contribution is expressed as: [7, 20, 38]

$$\sigma_C^{\mu:\alpha\beta}(0;\omega,-\omega) = \frac{2e^3}{S\omega^2} \sum_{k\in BZ} \sum_{lmn,\Omega=\pm\omega} f_{ln} \frac{v_{k,mn}^\mu v_{k,nl}^\alpha v_{k,lm}^\beta}{(E_{k,mn} - i\hbar/\tau)(E_{k,lm} - \hbar\Omega)} \quad (1)$$

here *k* labels the k-point in irreducible BZ and *l*, *m*, *n* label the band index, $v_{k,lm}^\mu$ is the matrix elements of velocity operator, $E_{k,ml} = E_{k,m} - E_{k,l}$ denotes the difference in band energies, $f_{ln} = f_l - f_n$ is the difference of occupations in $l^{th}$ and $n^{th}$ bands. $\omega$ is the magnitude of frequency of incident light. $\tau$ is the relaxation time for intra-band process. In principle, the relaxation time $\tau$ might depend on momentum, band indices, and frequencies of light, because of impurity scattering, electron-phonons couplings, many-body interactions, etc. However, several previous studies have showed that it is still reasonable to consider an average and constant relaxation time approximation in the calculations of BPVEs.[39-40] Hence, in this study we adopt the constant relaxation time approximation and take the default value $\hbar/\tau = 1\ meV$, i.e., $\tau \approx 0.6$ ps by default. This setting is rather conservative since the MnPSe$_3$ is supposed to be a clean crystal due to its excellent chemical stabilities. $S$ is the area of MnPSe$_3$ cell. Since the thickness of 2D monolayer is not well-defined, we do not average the photoconductance over volume but area. So, the $\sigma_C$ calculated in eq. 1 is related to conventional definition of $2^{nd}$ order photoconductances in 3D systems $\sigma_C^{3D}$ by $\sigma_C = L\sigma_C^{3D}$, in which $L$ is the effective thickness of 2D monolayer. And the unit of $\sigma_C$ in eq. 1 is nm·μA/V$^2$.

Figure 2 shows the calculated nonlinear photoconductance of BPVE. For MnPSe$_3$ with Néel vector along $z$-axis [Fig. 2(a)], $\sigma_C^{x;xx}$ exhibits several peaks in the visible spectrum with photon energy range from 1.6 eV to 3.6 eV. The first peak of $\sigma_C^{x;xx}$ occurs at 1.83 eV, slighter higher than the energy gap of ~1.5 eV. The peak value is −187 (nm·μA/V$^2$), which is comparable to the highest BPVE conductance previously reported in bilayer AFM systems.[7, 9] The strongest peak of $\sigma_C^{x;xx}$ is at 2.98 eV, with magnitude 4152 (nm·μA/V$^2$), which is ~10-folds larger than the previous proposals. On the other hand, $\sigma_C^{y;xx}$ vanishes at all photon energies, consistent with the $M_y$-symmetry since $\sigma_C^{y;xx}$ is proportional to $v_y \times v_x \times v_x$ which is odd in $M_y$-reflection. Therefore, the photocurrents induced by the linearly polarized light only propagate along the $x$-axis in this case.

Note that because of the injection mechanism of BPVEs in MnPSe$_3$, the magnitude of photoconductance is proportional to the value of relaxation time.[7, 20] The explicit dependence of BPVEs on relaxation time are shown in Fig. S1 in supporting information. Since the utilized relaxation time constants of 0.6ps is conservative, the estimations on the BPVEs should be conservative, too.

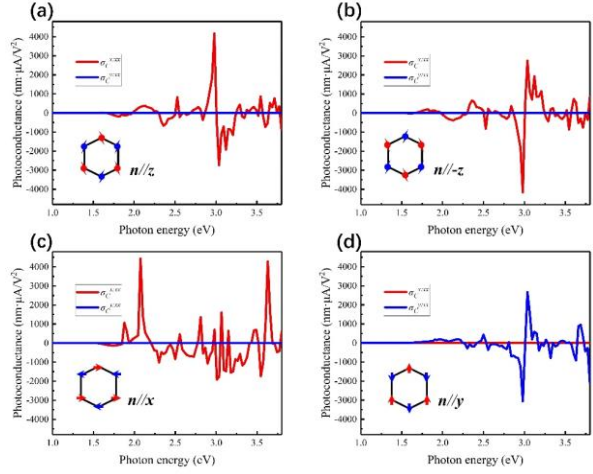

**Figure 2.** Photoconductance $\sigma_C^{x;xx}$ and $\sigma_C^{y;xx}$ for MnPSe$_3$ with Néel vector along (a) $z$-direction, (b) -$z$-direction, (c) $x$-direction, and (d) $y$-direction. The hexagon in the left-down part of each figure denotes the magnetic lattice formed by Mn$^{2+}$ ions, and the arrows on it denote the orientations of local magnetic moments. The directions of Néel vectors are marked aside the hexagons.

Then we reverse the Néel vector to −$z$-direction, and the $\sigma_C^{x;xx}$ displayed in Fig. 2(b) becomes the opposite of the case discussed before, indicating that the propagation directions of photocurrents are locked with the orientation of Néel vectors. This is in consistency with our previous discussions on $T$-symmetry i.e., the photocurrent propagation is related to $PT$-polarization, which is further controlled by the magnetizations. And this behavior suggests a promising route to read out the nonvolatile information stored as Néel vectors in MnPSe$_3$ via the BPVEs, which is essential for the 2D opto-spintronic memories.

Fig. 2(c) displays the photoconductance of MnPSe$_3$ with Néel vector along $x$-axis. Once again, due to the $M_y$-symmetry, $\sigma_C^{y;xx}$ is zero everywhere. The first peak of $\sigma_C^{x;xx}$ now occurs around 1.73 eV with value −100 (nm·μA/V$^2$), and the highest peak of $\sigma_C^{x;xx}$ in Fig. 1(c) is 4436 (nm·μA/V$^2$) at 2.02 eV. Comparing to the ±$z$-cases discussed before, the photocurrents induced by low-energy photons is smaller here. This can be explained by the valley structures show in Fig. 1(d). Since the valley-polarizations in $x$-magnetized system are much smaller, the low-energy photoconductance, which is dominated by the valley-polarization, should be weaker. Besides, the several high peaks in Fig. 2(c) have magnitudes comparable with ±$z$-magnetized cases, revealing that the high-energy photoconductance is free from the valley-related opto-electronic processes.

Fig. 2(d) displays the case with Néel vector along $y$-axis. $\sigma_C^{x;xx}$ becomes zero at all photon energies, thus the photocurrents generated by linearly polarized light run in $y$-axis solely. This perpendicular switching of photocurrents propagation is caused by the switching of mirror symmetries from $M_y$ to $M_xM_z$, which is brought about by the rotations of Néel vectors from $x/z$-axis to $y$-axis. The $\sigma_C^{x;xx}$ is odd in $M_y$ mirror operation, thus no photocurrents in $x$-axis are allowed. For $\sigma_C^{y;xx}$, it reaches the first peak at 1.97 eV with value 241 (nm·μA/V$^2$), and the highest peak value is −3059 (nm·μA/V$^2$) occurred at photon energy of 2.98 eV.

In addition, the 2D BPVEs of MnPSe$_3$ also generate pure spin-currents. We detect the spin-resolved photocurrents with vector potentials projected to spin tunnels, and compute the nonlinear photo-spin-conductance tensor $\sigma_{Sz}^{\mu;\alpha\beta}$, which is relevant for spin-currents (See supporting information for more details). The spin polarizations in photocurrents are dominate by the $z$-component in most cases, we thus focused on the photo-spin-conductance with $z$-polarization, which are displayed in Figure 3.

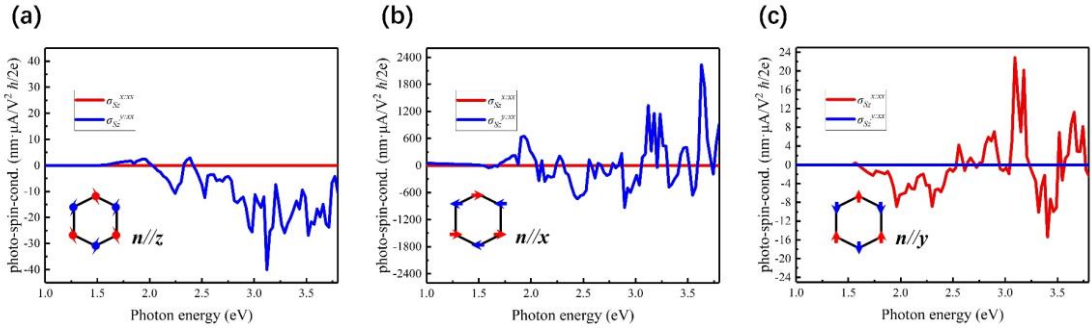

**Figure 3.** Photo-spin-conductance $\sigma_{Sz}^{x;xx}$ and $\sigma_{Sz}^{y;xx}$ for MnPSe3 with Néel vector along (a) $z$-direction, (b) $x$-direction, and (c) $y$-direction.

Figure 3(a) shows the photo-spin-conductance in system with Néel vectors along $z$-axis. Firstly, $\sigma_{Sz}^{x;xx}$ vanishes at all energies. $\sigma_{Sz}^{y;xx}$ acquires its first peak value 6 (nm·μA/V$^2$ $\hbar/2e$) around 1.99 eV. The highest peak value is -40 (nm·μA/V$^2$ $\hbar/2e$) occurred at photon energy of 3.12 eV. When the Néel vector rotate to $x$-axis, $\sigma_{Sz}^{x;xx}$ keeps zero values, but the $\sigma_{Sz}^{y;xx}$ is drastically enlarged [Fig. 3(b)]. The first peak of $\sigma_{Sz}^{y;xx}$ is -48 (nm·μA/V$^2$ $\hbar/2e$) at 1.61 eV. The strongest $\sigma_{Sz}^{y;xx}$ peak in Fig. 3(b) is

2237 (nm·μA/V$^2$ $\hbar$/2e) at 3.6 eV. For the MnPSe$_3$ with Néel vector in *y*-axis, $\sigma_{S_z}^{x;xx}$ presents finite magnitudes while $\sigma_{S_z}^{y;xx}$ vanishes [Fig. 3(c)]. And $\sigma_{S_z}^{x;xx}$ get the first peak at 1.75 eV with value -2 (nm·μA/V$^2$ $\hbar$/2e), and the largest $\sigma_{S_z}^{x;xx}$ takes place at 3.09 eV with value 23 (nm·μA/V$^2$ $\hbar$/2e).

Hence, the photo-spin-currents are modulable via rotating the Néel vectors, which can be explained by the switching of mirror symmetries. The photo-spin-conductance is related to the conventional photoconductance by the equation $\sigma_{S_z}=s_z\times\sigma_C$, and $s_z$ is odd about both $M_y$- and $M_xM_z$-reflections. Therefore, the allowed propagation directions of spin currents (with $s_z$-polarization) are always orthogonal to the charge currents, i.e., the spin currents found here are pure without the net movements of charges, which are desired for the low-consumption spintronics. Besides, it worths noticing that the photo-spin-conductance for cases with Néel vectors along *x*-axis is stronger than the other two cases by magnitude orders, indicating that the photon-to-spin conversions is switched from shift-dominated mechanism to injection-dominated mechanism along with the rotations of Néel vectors from *y*/*z*- to *x*-axis. This change in mechanism is further supported by the analysis of the dependence of photo-spin-conductance on relaxation time (see Fig. S2 in supporting information for more details). The injection-type spin-currents are intriguing and distinguished from the previous reports,[41-42] and the significantly large injection-spin-currents might open a new way for robust spin-currents generations via illuminations.

To further investigate the structure of BPVs in MnPSe$_3$, we plot the Brillouin zone (BZ) distribution of photoconductance in Figure 4. The photon energy is 2.98 eV, corresponding to the position of the strongest BPVE in Fig. 2. The most relevant states to the BPVE form several butterfly-like regions in BZ. and their contributions and positions are controlled by the orientations of Néel vectors, leading to the modulations on BPVE. Fig. 4(a) displays the BZ-distributions of $\sigma_C^{x;xx}$. The states in right part of BZ have positive crystal moments in *x*-direction and positively contribute to the $\sigma_C^{x;xx}$ (blue colored), while the states in left part of BZ are negatively related to the net $\sigma_C^{x;xx}$ (red colored). The topmost inset in Fig. 4(a) shows the case with Néel vectors along +*z*-direction. The most relevant states are located at the right part of BZ. The contributions from individual Bloch states reach as high as $4\times10^6$ (nm·μA/V$^2$), three orders larger than the net value of $\sigma_C^{x;xx}$, indicating the net BPVE is the residual effect of many counteracting contributions. It is thus hopeful to obtain ultra-large BPVE if one can optically excite the states with specific crystal momentums. When Néel vectors rotate to +$x$-direction [top right in Fig. 4(a)], the contributions from left and right parts of BZ are both clear. For Néel vectors in +$y$-direction [bottom right in Fig. 4(a)], the contribution regions move to the up part of BZ and become anti-symmetric to each other, in line with the zero-value of $\sigma_C^{x;xx}$ in this case. The lowest inset in Fig. 4(a) shows the case with Néel vectors in −$z$-direction. The contributions are dominated by the left part of BZ with negative values, which is exactly the opposite of the case displayed in the topmost inset. Further rotating the Néel vectors to −$x$- and −$y$-directions, we see the BZ-distributions of BPVE contributions are the opposite of +$x$- and +$y$-cases.

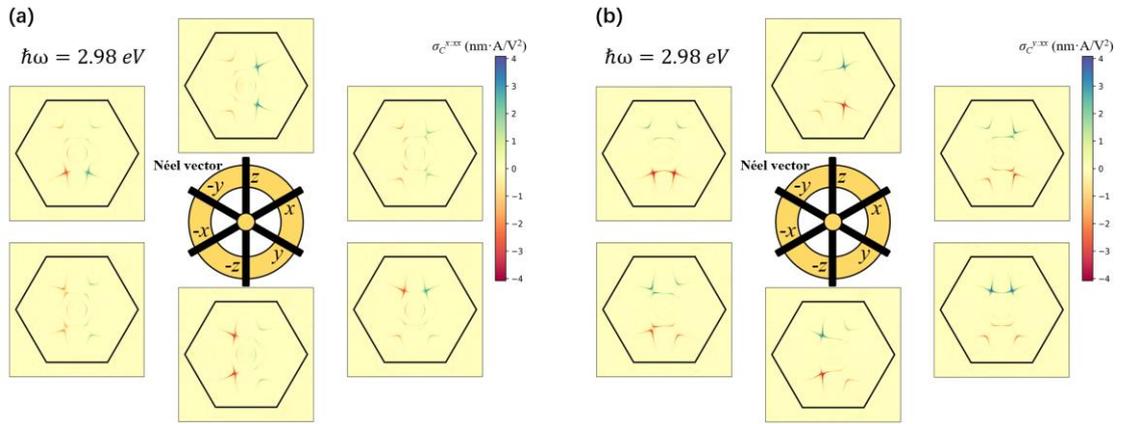

**Figure 4.** Photoconductances distributions on Brillouin zone that depend on the direction of Néel vector of MnPSe$_3$. (a) The BZ distributions of $\sigma_C^{x;xx}$. (b) The BZ distributions of $\sigma_C^{y;xx}$. The incident light is linearly polarized in $x$-axis, with photon energy of 2.98 eV. The distributions placed at top, top right, bottom right, bottom, bottom left, and up left correspond to the cases with Néel vectors along +$z$, +$x$, +$y$, −$z$, −$x$, and −$y$ directions. The hexagons are the irreducible Brillouin zone. For better illustrations, $\tau$ is set to 0.1 ps here.

Figure 4(b) displays the BZ-distributions of $\sigma_C^{y;xx}$. They have the same positions as the distributions of $\sigma_C^{x;xx}$ shown in Fig. 4(a), but the signs of contributions are different. The upper part of BZ corresponds to the Bloch states with positive crystal moments in $y$-direction, and the contributions in this region are blue, positively contributing to the net $\sigma_C^{y;xx}$. The contributions of lower part of BZ are in red color, exhibiting negative contributions to net $\sigma_C^{y;xx}$. For cases with Néel vectors along ±$z$- and ±$x$-directions, the contributions are anti-symmetric in the BZ, leading to the zero BPVEs. Only if the Néel vectors orient to ±y-directions, the contributions from upper and lower parts of BZ are not anti-symmetric and nonzero BPVEs are permitted. All these results are in line with former discussions on symmetries and agree with the integrated results displayed in Fig. 2.

To further understand the *PT*-BPVE in local coordinates and figure out how the *PT*-polarizations interact with the dynamics of photo-electrons, we detect the real-space resolved photoconductance with vector potentials projected to hopping tunnels (See supporting information for more details), which are the spaces between atoms as shown in Fig. 5(a). Below, we focused on the cases with Néel vectors along *z*-axis.

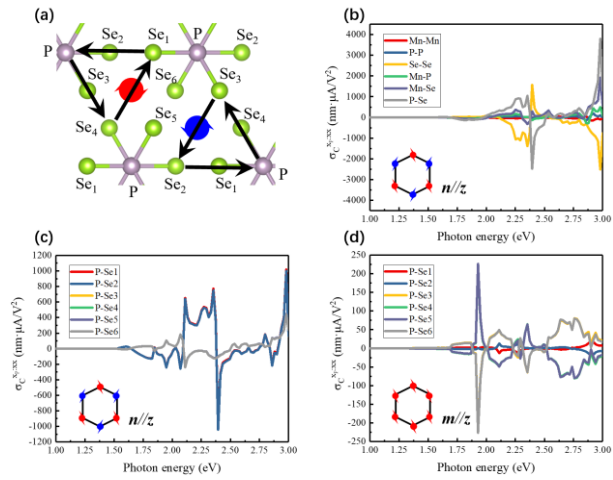

**Figure 5**. Bulk photoconductance projected to real-space. (a) Atomic positions and possible hopping tunnels in MnPSe$_3$. The arrows denote several typical hopping tunnels in $(P_2Se_6)^{4-}$. (b) $\sigma_C^{x;xx}$ in AFM MnPSe$_3$ projected to six types of tunnels. (c) $\sigma_C^{x;xx}$ in AFM MnPSe$_3$ projected to six types of P-Se tunnels. (d) $\sigma_C^{x;xx}$ in FM MnPSe$_3$ projected to six types of P-Se tunnels.

We firstly classify the photoconductance $\sigma_C^{x;xx}$ into six types according to the hopping tunnels including Mn-Mn, P-P, Se-Se, Mn-P, Mn-Se, and P-Se. The photocurrents are emitted by the electron-hole combinations happened in either one of these tunnels. And the summation of all these six types is exactly equal to the net BPVE displayed in Fig. 2(a).

Figure 5(b) shows the contributions from the six types. Comparing to the total $\sigma_C^{x;xx}$ shown in Fig. 2(a), it is clear that the P-Se tunnels always dominate the BPVEs with highest value of 3795 (nm·μA/V$^2$), while the contributions of Se-Se tunnels are comparable but they are negative and the strongest value is -2323 (nm·μA/V$^2$). The Mn-Se tunnels are the third largest contributors, and its highest peak is 1772 (nm·µA/V$^2$). Other tunnels contribute negligibly to the BPVEs. Therefore, the BPVEs in MnPSe$_3$ are mainly emitted by the hopping processes in the sublattice expanded by (P$_2$Se$_6$)$^{4-}$ prisms.

Fig. 5(c) shows the $\sigma_C^{x;xx}$ further projected to the six types of P-Se tunnels, including tunnels of P-Se$_1$, P-Se$_2$, P-Se$_3$, P-Se$_4$, P-Se$_5$, P-Se$_6$. The summation of all these six P-Se types is identical to the contributions of P-Se tunnels displayed in Fig. 5(b) with grey lines. The real-space orientations of tunnels of P-Se$_1$ and P-Se$_2$ are close to the $x$-axis [See Fig. 5(a)], and their contributions are nearly coincided, showing the ideal phase alignment for the optoelectrical processes in these two tunnels. And they constitute the most parts of photocurrents in the (P$_2$Se$_6$)$^{4-}$ prism sublattice with highest peak values of ~1000 (nm·µA/V$^2$). This ideal phase alignment is surprising since these two tunnels are not symmetric counterparts in AFM MnPSe$_3$. The contributions from the other four P-Se tunnels are moderate, and their strongest peak value is 441 (nm·µA/V$^2$). They coincide with each other due to the $M_y$-mirror symmetry.

Fig. 5(d) shows the $\sigma_C^{x;xx}$ projected to the six P-Se tunnels [as the case in Fig. 5(c)] for the MnPSe$_3$ with parallel magnetic moments along $z$-axis, that is, the ferromagnetic (FM) case. Although the P symmetry of FM MnPSe$_3$ enforces the zero value for total BPVEs, the local contributions are generally nonzero. The contributions from P-Se$_1$ and P-Se$_2$ tunnels are heavily suppressed in this case, and the highest peak value is only 21 (nm·µA/V$^2$), 100 times smaller than the AFM case displayed in Fig. 5(c). On the other hand, the contributions of the other four P-Se tunnels show peak values of 227 (nm·µA/V$^2$), which is comparable to the AFM case, especially in the low phonon energy range (<2 eV). Therefore, on the view of local photoelectrical responses, the role of AFM order in Mn-sublattices is two-folds: aligning the phases of photocurrents, and enhancing the photon-to-currents efficiencies in the direction of *PT*-polarization ($x$-direction in present case).

In summary, we found the 2D monolayer MnPSe$_3$ is promising to realize large and controllable *PT*-BPVEs. The 2$^{nd}$ order photoconductance and photo-spin-conductance exceed 4000 (nm·µA/V$^2$) and 2000 (nm·µA/V$^2$ ℏ/2e) in the visible spectrum. Both the propagations and the spin-polarizations of photocurrents are switchable via rotating the Néel vectors. These switching are enabled by the mirror symmetries, which depend on the magnetization. The intralayer AFM orders of Mn$^{2+}$ sublattice show essential roles in the 2D BPVEs, since they can stabilize the *PT*-polarizations and lead to the phase alignments of photocurrents in the sublattice of $(P_2Se_6)^{4-}$. Furthermore, we showed that the *PT*-BPVEs in MnPSe$_3$ are dominated by a small part of Bloch states in BZ, whose positions and contributions are intimately controlled by the Néel vectors. All these results shed light into the *PT*-BPVEs in 2D anti-ferromagnets, and indicate that the 2D monolayer MnPSe$_3$ is an ideal platform to achieve extraordinary *PT*-BPVEs, meriting future energy and spintronic applications.

**Computational Method.** The geometry optimization and electronic structure of MnPSe$_3$ in ground state are calculated within Vienna atomic simulation pack (VASP),[43] based on relativistic density functional theory (DFT). Projected augmentation plane wave basis (PAW) is utilized and the plane waves are cutoff at 400 eV. The exchange-correlation effects amongst electrons are captured via generalized gradient approximation (GGA) with the functional of Perdew-Burke-Ernzerh (PBE) form.[44] A vacuum layer of 20 Å is set to isolate the layers in nonperiodic direction to eliminate the unphysical interactions between neighboring slabs. Brillouin zone is sampled with a Γ-centered k-mesh 10×10×1 using Monkhorst-Pack scheme.[45] The geometry relaxation is carried out until the maximum force is smaller than 0.001 eV/Å. Energy convergent criteria for the self-consistent iterative calculations on the electronic structures is 10$^{-6}$ eV/atom.

Since the GGA functional usually underestimates the strong interactions between d-electrons of transition metal compounds (in our case, standard GGA leads to metal ground states of MnPSe$_3$, which is inconsistent with the experiments), we employed the GGA + U$_{eff}$ method with U$_{eff}$ = 3.0 eV for d-electrons of Mn. This value of U$_{eff}$ leads to the correct magnetic ground states and is in line with former studies on the MnPS$_3$.[32]

For the calculations of BPVE based on eq. 1 and related expressions, we construct the effective real-space tight-binding Hamiltonian via projecting the Bloch states obtained from relativistic DFT into the Hilbert space expanded by Wannier orbitals using WANNIER90,[46-47] then the k-space Hamiltonians on general k-points are interpolated by solving the eigen-equations. The Wannier orbital basis include the 3d and 4s orbitals for Mn ions, 3s and 3p orbitals for P ions, and 4s and 4p orbitals for Se ions. It is found that the effective tight-binding Hamiltonian can reproduce the electronic states predicted by relativistic DFT and the band structures of the effective model and the DFT coincide everywhere in the relevant energy window (see Figure S3 in Supporting information for more details), revealing the efficiency of the basis transformation. The k-grid utilized for the Brillouin zone integration in equation 1 is 800×800×1. A denser k-grid 1600×1600×1 is used to test the convergence of k-mesh sampling and we find the difference < 5%.

**Notes**

The authors declare no competing financial interest.

# ACKNOWLEDGMENT

This work was supported by Natural Science Foundation of Shandong (Grant No. ZR2022QA019), National Natural Science Foundation of China (Grant Nos. 12074221, 52171181, 52002222, 51472150, 2021-869, 11904204).

**Supporting information available**: Extracting out the expression of BPVEs from potentials; Expressions of spin and real-space projected BPVEs.